\documentclass[12pt]{article}
\usepackage{amsfonts}
\usepackage{amsmath, amsthm}
\usepackage{epsfig}
\usepackage{algorithm}
\usepackage{algorithmic}
\usepackage{subcaption}
\usepackage{epic}
\usepackage[a4paper, bottom = 1.50cm,
        ]{geometry}
\usepackage{enumitem}
\usepackage{makecell} 
\usepackage{amsmath,verbatim,color,amssymb,epsfig}
\usepackage{bm}
\usepackage{ragged2e}
\usepackage{amsfonts}
\usepackage{epsfig}
\usepackage{changepage}
\usepackage{multirow}
\usepackage{graphicx}
\usepackage{array}
\usepackage[table]{xcolor}
\definecolor{Gray}{gray}{.80}
\usepackage{lscape}
\usepackage[round,semicolon,authoryear]{natbib}
\usepackage[normalem]{ulem}
\usepackage[colorlinks=true,urlcolor=blue,citecolor=purple,linkcolor=blue,bookmarks=true]{hyperref}
\usepackage{threeparttable}
\usepackage{caption}
\usepackage{pdfsync}
\usepackage{array}

\usepackage{booktabs}  
\usepackage{cleveref} 
\usepackage{caption}  
\usepackage{tabularx}   
\usepackage{adjustbox} 
\usepackage{longtable}  

\begin{document}

\baselineskip=12pt
\begin{center}
{\Large \bf BE-BOIN: A Dose Optimization Design Accommodating Backfill and Late-Onset Toxicity}
   
\end{center}
\vspace{1mm}
\begin{center}
{\bf Kai Chen$^{1,2}$, Yixuan Zhao$^{2}$, Kentaro Takeda$^{3}$,  Ying Yuan$^{1,*}$}
\end{center}

\noindent$^{1}$Department of Biostatistics, The University of Texas MD Anderson Cancer Center, \\ Houston, TX \\
$^{2}$Department of Biostatistics and Data Science, The University of Texas Health Science \\ Center, Houston, TX\\
$^{3}$Astellas Pharma Global Development Inc., Northbrook, IL, USA \\
$^*$Author for correspondence: yyuan@mdanderson.org 
\vspace{2mm}

\baselineskip=24pt

\begin{center}
\noindent \emph{\textbf{Abstract}}
\end{center}

The US Food and Drug Administration (FDA) launched Project Optimus and issued guidance to reform dose-finding and selection trials, shifting the paradigm from identifying the maximum tolerable dose (MTD) to determining the optimal biological dose (OBD), which optimizes the risk and benefit of treatments. The FDA's guidance emphasizes the importance of collecting sufficient toxicity and efficacy data across multiple doses and considering late-onset cumulative toxicity that often results in tolerability issues. To address these challenges, we propose the BE-BOIN (Backfill time-to-Event Bayesian Optimal INterval) design, which allows backfilling patients into safe and effective doses during dose escalation and accommodates late-onset toxicities. BE-BOIN enables the collection of additional safety and efficacy data to enhance the accuracy and reliability of OBD selection and supports real-time dose decisions for new patients. Our simulation studies show that BE-BOIN accurately identifies the MTD and OBD while significantly reducing trial duration.

\bigskip
\noindent \emph{\textbf{keywords}}: Project Optimus, Dose finding, Backfill, Late-onset toxicity,  BOIN design.

\clearpage
\section{Introduction}
%

The FDA launched Project Optimus and issued guidance to shift the dose finding and selection paradigm from finding the maximum tolerable dose (MTD) to identifying the optimal biological dose (OBD), which aims to balance efficacy and safety to optimize the risk-benefit profile of investigational drugs. Accordingly, the focus of dose-finding trials should shift from the conventional approach of escalating the dose to the highest tolerable level to comparing multiple dose levels in terms of risk and benefit, as emphasized in the FDA's guidance. Achieving this goal requires collecting a reasonable amount of data across multiple doses. \citet{Yuan2024DOreview} and \citet{Huang23} provide a comprehensive review of these methods,  including efficacy-integrated designs, two-stage designs, and seamless phase 2-3 designs.  
In particular, the FDA guidance highlights backfilling as a useful strategy to collecting more data across multiple doses, stating: ``It may be useful to evaluate additional dose-level cohorts or add more patients to existing dose-level cohorts (i.e., backfill cohorts) in the dose-finding trial for dosages being considered for further development. This would provide additional clinical data to allow for further assessment of safety and activity prior to initiating a trial to compare multiple dosages".  

A number of methods have been proposed to incorporate backfilling into phase I trials, which involve concurrently assigning patients to doses previously deemed safe and showing preliminary efficacy during escalation. \citet{BF2021} proposed randomizing patients to backfill doses based on hypothesis testing. \citet{BFCRM} studied the effect of backfill on the estimation of the MTD and the trial duration based on Bayesian logistic regression model. \citet{BFQoL} proposed a Bayesian design that uses patient-reported outcomes to determine doses for backfill.  \citet{Pin2024} considered using response-adaptive randomization to allocate patients among backfill doses. \citet{BFBOIN} proposed the BF-BOIN design, which seamlessly integrates backfill into the BOIN framework with formal statistical criteria for opening and closing backfill cohorts and resolving potential data conflicts between dose escalation and backfill. Compared to other backfill designs, such as \citet{BF2021} and \citet{BFCRM}, which require complex model fitting and estimation, a key advantage of BF-BOIN is its ability to preserve the hallmark simplicity and robust performance of BOIN. The BF-BOIN design has already been implemented in a number of clinical trials (see, for example, ClinicalTrials.gov identifiers NCT05902988, NCT06908434, and NCT06463340). Subsequently, \citet{BARD} proposed the BARD design that enables the pooling of dose escalation and backfill patients with subsequent randomized patients to efficiently inform the selection of the OBD. 

Another important issue that motivated Project Optimus and is emphasized in the FDA's guidance is the late-onset toxicity and tolerability issue. The toxicity of targeted therapies is often late-onset \citep{Postel2011, Weber2015, Kanjanapan2019}. For example,  \citet{Postel2011} reported that in 36 clinical trials of molecularly targeted agents, 57\% of grade 3–4 toxicities occurred after the first treatment cycle. Additionally, as noted by \citet{shah2021drug} and \citet{Zirkelbach2022}, acute DLT is seldom observed in many novel targeted agents. Instead, as these agents are often administered over multiple cycles, they frequently result in cumulative late-onset toxicities, causing tolerability issues that lead to dose interruption, reduction, and discontinuation. To capture these late-onset toxicities, it is imperative to use a longer toxicity observation window. This poses tremendous logistical and operational challenges for sequential dose-finding designs, which typically require fully observed data before making dosing decisions for the next cohort of patients.

Several designs have been proposed to address late-onset toxicity, including model-based designs such as TITE-CRM \citep{TITECRM}, DA-CRM \citep{DACRM}, and EM-CRM \citep{EMCRM}, as well as model-assisted designs like TITE-BOIN \citep{TITEBOIN}. These designs account for patients who are still within the toxicity assessment window (referred to as "pending patients") by leveraging their follow-up data to enable real-time dose decisions for new cohorts. For instance, TITE-CRM incorporates pending patients by weighting their likelihood based on the fraction of their toxicity window that has been completed. The TITE-BOIN design imputes pending toxicity data for pending patients using their follow-up data and combines these imputed data with observed data to make real-time decisions. TITE-BOIN has been successfully applied in various trials, including those for targeted therapies (ClinicalTrials.gov ID: NCT04388852), immunotherapies (NCT03956680), radiotherapies (NCT06617169), and chemotherapies (NCT05620654).

In this article, we propose BE-BOIN (Backfill TITE-BOIN), an extension of BF-BOIN that accommodates late-onset toxicity using the TITE-BOIN approach. We provide the rules for opening and closing backfill, as well as strategies for addressing potential conflicts when backfill doses exhibit higher toxicity rates than the current dose-escalation dose. A unique challenge in this setting is incorporating pending toxicity outcomes from both dose-escalation and backfill patients into decision making. To address this, we adopt the imputation approach from TITE-BOIN to handle pending data effectively. Simulation studies demonstrate that BE-BOIN exhibits desirable operating characteristics, including shorter trial durations compared to BF-BOIN, the ability to treat more patients than TITE-BOIN, and often improved accuracy in identifying the MTD and OBD. As a byproduct, our results also demonstrate that the BF-BOIN approach of ignoring pending backfilling patients has a negligible impact on the operating characteristics of the design, addressing a common question raised about BF-BOIN. 

The remainder of this article is organized as follows. Section 2 introduces the time-to-event model and the BE-BOIN design. Section 3 presents the simulation study, and the article concludes with a discussion in section 4.

\section{Methods}
\subsection{Handling pending data}
A key challenge in trials involving late-onset toxicity and backfill is managing pending data, which hinders real-time dose-decision-making for the next cohort of patients. Pending data arise not only from patients in dose escalation but also from those in backfill. In the following sections, we first discuss methods for handling pending data, followed by the design decision rules. 

Assume there are $J $ dose levels, and let $p_j$ denote the dose-limiting toxicity (DLT) rate of dose $j$, where $ j = 1, 2, \ldots, J.$  Suppose at an interim decision time, $n_j$ patients have received dose  $j$.  Let $x_i$ denote the DLT indicator for patient $i$, $i=1, \cdots, n_j$,  with $x_i=1$ indicating DLT and $x_i=0$ indicating no DLT. Let $O_j$ represent the group of patients who have completed the DLT assessment, such that $x_i$ is known, and $M_j$ represent the group of patients whose DLT assessments are still pending, where $x_i$ is unknown.  We estimate $p_j$ as:
$$
 \hat{p}_j = \frac{\sum_{i \in O_j} x_i + \sum_{i \in M_j} x_i}{n_j}.
$$
The challenge in calculating $p_j$ is that $x_i \in M_j$ is unobserved. 

Let $u_i$ denote the time to DLT, and $t_i$ denote the follow-up time for patient $i$ at the interim decision time.  Following the TITE-BOIN \citep{TITEBOIN},  and assuming $u_i$ follows a uniform distribution over the DLT assessment window $(0,\tau)$ for patients with $x_i=1$, the unobserved $x_i \in M_j$, with a follow-up time $t_i \le \tau$, can be imputed using its expectation:
\begin{align*}
    \hat{x}_i &= E(x_i | u_i>t_i) = \Pr(x_i =1|u_i>t_i ) \\
    &= \frac{\Pr(u_i>t_i |x_i=1)\Pr(x_i=1)}{\Pr(u_i>t_i |x_i=1)\Pr(x_i=1) + \Pr(u_i>t_i |x_i=0)\Pr(x_i=0)} \\
    &=\dfrac{(1-\dfrac{t_i}{\tau})p_j}{(1-\dfrac{t_i}{\tau})p_j + (1-p_j)}
    \approx \dfrac{(1-\dfrac{t_i}{\tau})p_j}{1-p_j}.
\end{align*}
The rationale for the approximation is that $p_j$ and consequently (1-${t_i}/{\tau})p_j$ are often small relative to $1-p_j$.
After the imputation, $\hat{p}_{j}$ is given as: 
\begin{equation}
\hat{p}_j = \frac{\Tilde{y}_j + \dfrac{p_j}{1-p_j}(m_j - T_j)}{n_j},
\label{p_j}
\end{equation}
where $\Tilde{y}_j = \sum_{i \in O_j}x_i$ is the number of observed DLTs, $m_j$ is the total number of pending patients, and $T_j=\sum_{i \in M_j} {t_i}/{\tau}$ is the standardized total follow-up times for the $m_j$ pending patients at dose $j$. To calculate $\hat{p}_{j}$, the unknown $p_j$ can be replaced by its Bayesian posterior mean estimate, $\Tilde{p}_j=(\alpha + \Tilde{y}_j)/(\alpha + \beta + n_j -m_j)$, obtained from a beta-binomial model with the prior $ p_j \thicksim \text{Beta}(0.5\phi ,1-0.5\phi)$. This vague prior corresponds to an effective prior sample size of 1, with the prior mean equal to $0.5\phi$, where $\phi$ is the target DLT rate.


Following the same approach, we can also calculate the pooled toxicity rate $\hat{q}_{(b,k)}$ across doses from $b$ to a higher $k$, where $1\le b < k \le J$,
\begin{equation}
    \hat{q}_{(b,k)} = \frac{\sum_{j = b}^k \Tilde{y}_j + \sum_{j = b}^k\dfrac{\tilde{p}_j}{1-\tilde{p}_j}(m_j -T_j) }{\sum_{j = b}^k n_j}.
    \label{p_pool}
\end{equation} 
$\hat{q}_{(b,k)}$ will be utilized to select eligible doses for backfilling and to reconcile data conflicts arising from violations of toxicity monotonicity, as discussed in the next section. 

\subsection{Trial design}
\begin{figure}
\centering
\hspace*{-1.5cm}  
    \includegraphics[width=1.2\textwidth]{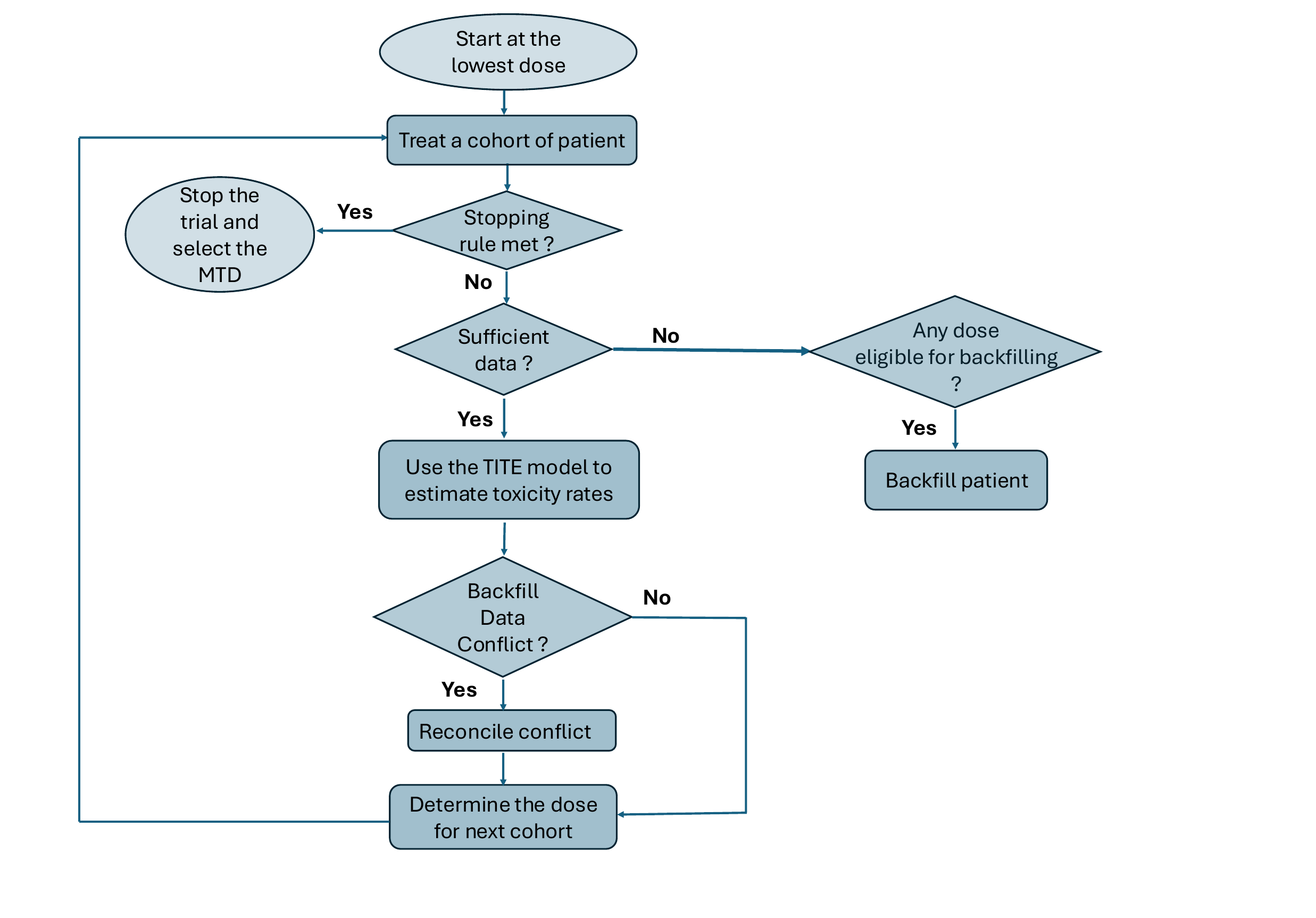}
    \caption{Flowchart of the BE-BOIN design.  Note that when the trial stopped for MTD selection, all patients have completed their DLT assessments, and cumulative data from all treated patients are used for the final analysis.}
    \label{flowchart}
\end{figure}

The BE-BOIN design consists of two interconnected  components: dose escalation (DE) and backfill. In the DE component, BOIN design rules are applied to explore the dose space and identify the MTD by the end of the trial. The backfill component is opened when a dose is deemed safe and demonstrates preliminary activity based on data from DE. Once opened, additional patients are concurrently treated at one of these doses while DE continues. At both interim and final analyses, cumulative data from DE and BF are combined to guide decision-making. Figure \ref{flowchart} presents the flowchart of the BE-BOIN design. In the following sections, we describe DE, followed by backfill.

\subsubsection{Dose escalation}
In DE, patients are treated in cohorts. The most commonly used cohort size is three, but it can be adjusted based on the characteristics of the trial. Suppose at a dose decision time for the next cohort, and current cohort is treated at dose $c$ in DE. Let $\lambda_e $ and $\lambda_d$ denote escalation and de-escalation boundaries of the BOIN design \citep{BOIN2015}. BE-BOIN determines dose escalation or de-escalation decisions as follows:

\begin{enumerate}
\item Treat the first cohort of patients at the lowest dose or at pre-specified doses.
\item To assign a dose to the next cohort of patients, calculate $\hat{p}_c$ using equation (\ref{p_j}), 
    \begin{itemize}
    \item[] a. If $\hat{p}_c \le \lambda_e$, escalate to dose $c+1$;
    \item[] b. If $\hat{p}_c > \lambda_d$, de-escalate to dose $c-1$;
    \item[] c. Otherwise: retain the current dose $c$.  
    \end{itemize}
\item Repeat step 2 until the maximum sample size $N$ is reached.
\end{enumerate}
A rule of thumb for determining the maximum sample size $N$ in step 3 is to set it as $6\times J$ \citep{Yuanbook2022}. For instance, in a phase I trial with five dose levels, $N$ would be set to 30. This value can be further calibrated to ensure desirable operating characteristics. 

As a tradeoff for the ability to make decisions in the presence of pending patients, the described DE approach is associated with greater uncertainty compared to fully staggered designs, as patients with pending DLT outcomes might later exhibit a DLT after a decision has already been made \citep{Biard2024}. This uncertainty increases with the number of pending patients at the time of dose decision-making. To mitigate this concern and ensure decisions are made based on adequate data and follow-up time, we implement two accrual suspension rules \citep{TITEBOIN2}: 
\begin{quote}
    	\textbf{Rule 1}: Suspend accrual to wait for more DLT data if fewer than $51$\% of patients at the current dose have their DLT outcome observed, unless the observed DLT rate has already exceeded  $\lambda_d$. 
\end{quote}
\begin{quote}
	\textbf{Rule 2}: Suspend accrual to allow for longer follow-up of pending patients if the data recommends dose escalation and the shortest follow-up time for pending patients at the current dose is less than $25$\% of the DLT assessment window.
\end{quote}
By incorporating these rules, the design prioritizes patient safety while allowing for more reliable dose decision with pending patients. The cutoffs of 51\% in Rule 1 and 25\% in Rule 2 are recommended default values, but they are adjustable. 

One prominent advantage of the TITE-BOIN framework is that the dose escalation/de-escalation rules described above can be pre-tabulated prior to trial initiation (see Supplementary material S1 for an example). Consequently, dose escalation and de-escalation decisions can be easily made by referring to the decision table, greatly simplifying its implementation. This contrasts with other methods, such as TITE-CRM, which require repeated model fitting and estimation.

\subsubsection{Backfill patients}
In the DE component, if suspension rules are triggered due to a high number of pending patients or insufficient follow-up time, the accrual of new patients to the DE should be paused. 
In this case, newly enrolled patients can be backfilled to safe, lower doses that demonstrate reasonable efficacy. This backfilling approach not only facilitates the collection of additional patient data across multiple doses for dose optimization but also enhances patient access to the trial, ensuring the continuity of accrual and improving overall trial efficiency. This feature is particularly advantageous in scenarios with late-onset toxicity and a long DLT window, where newly enrolled patients are less likely to be assigned to DE in a timely manner. 

To backfill patients, only doses showing both efficacy and safety are eligible for backfilling. Following \citet{BFBOIN}, given the current DE dose $c$, a lower dose $b$ $(b<c)$ is considered eligible for backfilling if it meets the following three criteria:
\begin{itemize}
\item  Safety: $\hat{p}_{b} \le \lambda_d$, or $\hat{p}_{b} > \lambda_d$ but the pooled estimate $\hat{q}_{(b, b+1)} \le \lambda_d$.
\item Efficacy: At least one response is observed at or below $b$.
\item Sample size: The total cumulative number of patients at dose $b$ (denoted as $n_b$, including patients from DE and backfill) satisfy $n_b \le n_{\text{cap}}$. 
\end{itemize}
In the safety criteria, the second condition (i.e., $\hat{p}_{b} > \lambda_d$ but the pooled estimate $\hat{q}_{(b, b+1)} \le \lambda_d$) accounts for cases where additional data from backfill patients suggest that dose $b$ might be overly toxic, exceeding the BOIN de-escalation boundary $\lambda_d$. However, the observed high DLT rate may occur sporadically due to the small sample size, as $b$ is a lower dose and, therefore, should be safer than $c$. To reduce the likelihood of accidentally excluding a dose for backfill, the pooled estimate is used to determine whether the high DLT rate observed at $b$ is reliable. As noted by \citet{BFBOIN}, the pooled estimate serves as an approximation to the isotonic estimate, offering nearly identical performance while being much simpler. In the efficacy criteria, the definition of response is trial-specific, ensuring that the dose for backfill is likely to provide therapeutic benefit to patients. Examples of response include objective response or biomarker response.


The above three criteria should be evaluated whenever a backfilling decision is being made. If the criteria are not met, backfilling should be paused for $b$. During the trial, it is possible for more than one dose to meet the backfill criteria. In such cases, different strategies can be employed to allocate patients. Patients can be randomized to all or a subset of backfill-eligible doses, or they can be assigned to the highest backfill-eligible dose, assuming that the efficacy of a higher dose is not lower than that of a lower dose.

The backfill eligibility criteria are structurally similar to those of BF-BOIN \citep{BFBOIN}, but they differ in the calculation of $\hat{p}_b$ and $\hat{q}_{(b, b+1)}$. Here, the calculations incorporate pending patient data, whereas BF-BOIN uses only observed data. In BF-BOIN, DE is fully staggered before cohorts, so there are no pending patients in DE. However, backfill patients are enrolled without staggering, which means pending patients may be present when evaluating the backfill eligibility criteria. A common question we received regarding BF-BOIN is how ignoring pending backfill patients impacts operating characteristics. BE-BOIN provides a solution to this issue, which will be discussed in detail in the Simulation section.

\subsubsection{Incorporating backfill data into DE}
Backfilling provides additional data on doses previously cleared for safety by DE, which are based on a small number of patients (e.g., 3 patients). A challenge arises if additional patient data from backfilling resulting in a lower dose $b$ having a higher toxicity rate than that of $c$. For example, if 0 out of 3 patients at $c$ experience DLT, suggesting an escalation to $c+1$, this decision may conflict with data from a dose $b$, where 2 out of 5 patients experiences DLT——indicating a need to de-escalate at $b$. In this scenario, the additional DLTs at $b$ are from backfilled patients (e.g., two backfilled patients at $b$ experienced DLTs, while none of the three patients at $b$ from DE experienced DLT). Table \ref{Dataconflict} summarizes potential data conflicts, defined as a situation where cumulative data at $b$ suggests a more conservative decision than the data at $c$. 

When a data conflict occurs, it may be due to the randomness associated with the small sample size at  $b$, or $b$ may truly be toxic. In the latter case,  some adjustment is needed for DE, noting that $c$ is higher and thus should be more toxic than $b$,  to ensure patient safety.  Following a similar approach proposed by \citet{BFBOIN}, we pool data across doses to obtain a more reliable estimate of toxicity at $b$, facilitating such decision.   

Specifically, let $b^*$ be the backfill dose (or the lowest dose when multiple doses are backfilled) whose data conflict with the data at the current dose $c$ of DE. We calculate the pooled toxicity estimate $\hat{q}_{(b^*,c)}$ from $b^*$ to $c$ using equation (\ref{p_pool}). The dose escalation/de-escalation rule in DE is adjusted as:
\begin{enumerate}
    \item If $\hat{q}_{(b^*, c)} \le \lambda_e$, escalate to higher dose $c+1$.
    \item If $ \lambda_e < \hat{q}_{(b^*, c)} \le \lambda_d$, stay at the current dose $c$.
    \item If $\hat{q}_{(b^*, c)} > \lambda_d$, de-escalate to the highest dose $k$ that is deemed safe with $\hat{q}_{(b^*,k)} \le \lambda_d$, where $b^* \le k <c$. If such a dose does not exist, de-escalate to dose $b^*-1$.
\end{enumerate}
Of note, if no data conflict is observed, no adjustment to DE is needed. One important advantage of using the pooled estimate is that the TITE-BOIN decision table provided in Supplementary material S1 can still be used to make decisions of dose escalation or de-escalation. 

Again, the approach of identifying and resolving data conflict is similar to those of \citet{BFBOIN}, but they differ in the calculation of $\hat{q}_{(b^*, c)}$. Here, the calculations incorporate pending patient data, whereas BF-BOIN uses only observed data. 


\subsubsection{Determination of MTD and OBD}
Upon the completion of DE and backfill, the MTD is selected based on the pooled data from DE and backfill patients. In BE-BOIN, we use the same dose selection rule as the BOIN: the MTD is selected as the dose whose DLT rate estimate, based on isotonic regression, is closest to the prespecified target $\phi$. However, as noted by \citet{BOIN2015}, DE and final MTD selection are two independent components, and BOIN is not restricted to isotonic regression. Any statistical model, such as a logistic model, can be fitted to the final data to provide the estimate of $p_j$ for MTD selection.

After the identification of the MTD, a multiple dose randomization phase is often added to collect additional efficacy and toxicity data to better characterize the OBD. The doses for randomization are selected based on the totality of toxicity, efficacy, and pharmacokinetic and pharmacodynamic data from both DE and backfill patients. Often, the MTD ($d_{high}$) and one lower dose ($d_{low}$) are selected, and patients can be equally or adaptively randomized to these doses. \citet{Yang24MERIT} suggested that a sample size of 20-40 patients per dose arm often leads to reasonable accuracy in identifying the OBD.

At the conclusion of the multiple-dose randomization phase, various methods can be employed to determine the OBD. We adopt the utility-based method, a highly flexible and scalable approach \citep{BOIN12, UBOIN, UMET}, to select the OBD.

The utility-based approach integrates efficacy and toxicity data into a utility function to assess the risk-benefit trade-off. A higher utility score reflects a more favorable balance between efficacy and safety, with the OBD identified as the safe dose yielding the highest score. For binary efficacy and toxicity endpoints, each patient falls into one of four outcomes: (efficacy, no toxicity), (efficacy, toxicity), (no efficacy, toxicity), or (no efficacy, no toxicity). Clinicians assign utility scores ($u_1, u_2, u_3, u_4$) ranging from 100 (most favorable) to 0 (least favorable) based on outcome desirability. Let $(\pi_{j1}, \pi_{j2}, \pi_{j3}, \pi_{j4})$ be the probabilities of these four outcomes at dose $j$. Using a Dirichlet-multinomial model, we assume a prior distribution:
$$
(\pi_{j1}, \pi_{j2}, \pi_{j3}, \pi_{j4}) \sim \text{Dirichlet}(\eta_1, \eta_2, \eta_3,\eta_4),
$$
where  $\eta_1 = \eta_2 = \eta_3 = \eta_4 = 0.25$ , corresponding to an effective prior sample size of one patient. Given the observed event counts  $D_n = (n_{j1}, n_{j2}, n_{j3}, n_{j4})$ for each outcome, the conjugate posterior distribution is:
$$
(\pi_{j1}, \pi_{j2}, \pi_{j3}, \pi_{j4})|D_n \sim \text{Dirichlet}(\eta_1+n_{j1}, \eta_2+n_{j2},\eta_3+n_{j3},\eta_4+n_{j4}).$$
The posterior means $\hat{\pi}_{j1}$ to $\hat{\pi}_{j4}$ estimate the utility for dose $j$ as:
$$\hat{U}_j = \sum_{k=1}^{4} u_k \hat{\pi}_{jk}.$$
The OBD is determined by comparing $\hat{U}_{d_{high}}$ and $\hat{U}_{d_{low}}$, selecting the dose with the higher utility score.

\section{Simulation study}
\subsection{Simulation setting}
We conducted a simulation study to evaluate the performance of the proposed BE-BOIN design. The target DLT rate $\phi$ was set at 0.25. Five dose levels were considered, with a total sample size $N$ of 30 and a cohort size of 3. The sample size for backfill is capped at $n_{\text{cap}}=12$ at a dose. The DLT assessment window was set at three months, with an accrual rate of two patients per month. Statistically, the operating characteristics of a design are determined by the ratio of the DLT assessment window to the inter-patient arrival time (i.e., the reciprocal of the accrual rate), referred to as the A/I ratio, rather than the DLT window alone. For example, a one-month DLT assessment window with an accrual rate of six patients per month (fast accrual) yields the same operating characteristics as a three-month DLT window with an accrual rate of two patients per month (late-onset setting). The time to DLT event was sampled from Weibull distribution, with 50\% of DLTs occurring in the latter half of the assessment window. 

\begin{figure}
\centering
\hspace*{-1.5cm}  \includegraphics[width=1.2\textwidth]{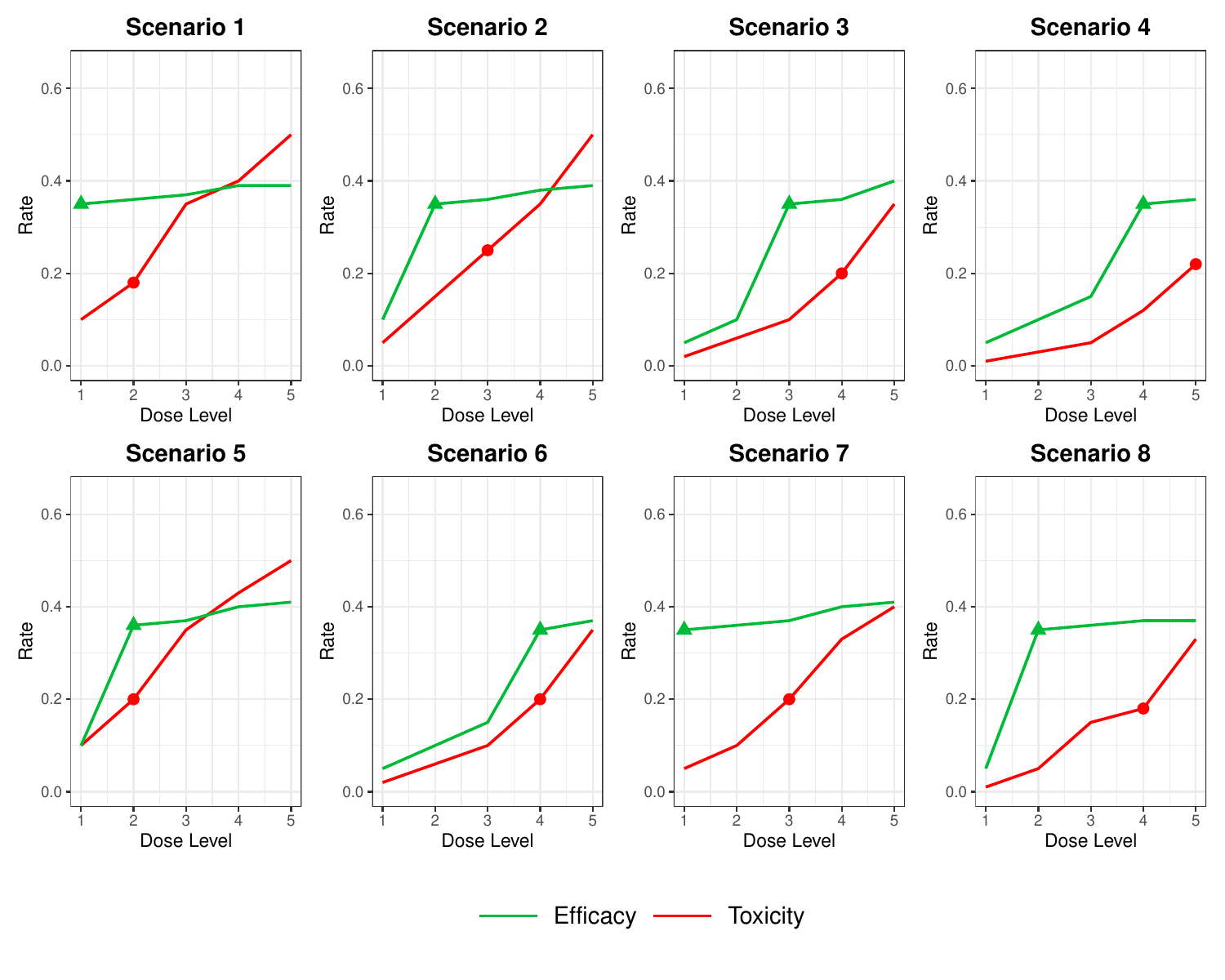}
   \caption{Plot of eight scenarios with different location of the maximum tolerated dose (MTD) and optimal biological dose (OBD), with MTD marked by circles and OBD indicated by triangles.}
   \label{scen}
\end{figure}

After identifying the MTD in stage I, the trial proceeded to a multiple-dose randomized phase for dose optimization in stage II. Two doses, including MTD and the dose one level lower, were selected, with each receiving 20 additional randomly assigned patients. The utility scores $(u_1, u_2, u_3, u_4)$ were set as (100, 60, 40, 0). 
Simulations were conducted 10,000 times for each of the eight scenarios (see Supplementary material S2), representing various dose-toxicity and dose-response relationships, as depicted in Figure \ref{scen}. Specifically, in scenarios 1-4, the OBD was located one dose level below the MTD. In the scenarios 5-6, the OBD coincided with the MTD, and in scenarios 7-8, OBD was located two dose level below the MTD. 

We compared the BE-BOIN design to the TITE-BOIN design, which does not incorporate backfill, and the BF-BOIN design, which does not support real-time decision-making. Additionally, we compared BE-BOIN to the model-based backfill TITE-CRM, denoted as BE-CRM \citep{BFCRM}. The BE-CRM model setting is present in the Supplementary material S3.  To ensure a fair comparison and facilitate the interpretation of the results, the same accrual suspension rules and backfilling timing were used in BE-CRM and BE-BOIN.

Additionally, we compared BE-BOIN to BF-BOIN in the absence of late-onset toxicity by setting the DLT window to one month and the accrual rate of three patients per month. The DE cohorts are staggered to ensure that no pending data from DE patients occurs. The objective is to determine whether ignoring pending data from backfill patients, as done in BF-BOIN, impacts its operating characteristics compared to including pending data. We also conducted  sensitivity analysis to evaluate the robustness of BE-BOIN. These analyses addressed the uniform time-to-event assumption, the impact of varying the A/I ratio from 3 to 9, and the effects of a high incidence of DLTs occurring later in the assessment window, such as over 50\%.

\subsection{Simulation Results}
Table \ref{taboc} summarizes the simulation results comparing TITE-BOIN, BF-BOIN, BE-BOIN, and BE-CRM. After stage I, BE-BOIN achieves a robust accuracy of correct MTD selection(\%), averaging 55.4\% (range: 46.2–67.9), which exceeds TITE-BOIN’s 52.6\% (range: 42.7–68.3) by an average of 2.7\%. Compared to BF-BOIN, which leads with an average MTD selection of 56.1\% (range: 44.5–68.8), BE-BOIN trails by only 0.7 percentage points. Across most scenarios, BE-BOIN outperforms BE-CRM, with an average improvement of 3.4\% over BE-CRM’s 52.0\% (range: 37.6–86.9). However, in scenario 4—where the MTD is the highest dose—BE-CRM achieves superior accuracy (86.9\% vs. 67.9\%), attributed to its tendency of aggressive escalation. On average, BE-CRM overdosed 3.9 more patients than BE-BOIN. For example, in scenarios 6 and 7, BE-BOIN overdosed 9.7 and 13.8 patients, respectively, while BE-CRM overdosed 16.5 and 19.4 patients.

After stage II, BE-BOIN maintains high accuracy in selecting the correct OBD across scenarios 1-6, averaging 63.9\% (range: 47.7–76.6), which improves upon TITE-BOIN (61.8\%, range: 45.9–73.1) by an average of 2.1 percentage. Compared to BF-BOIN, which achieves 64.6\% (range: 49.4–77.4), BE-BOIN remains highly comparable with only a 0.7 percentage point difference. Notably, BE-BOIN outperforms BE-CRM in scenarios 1–4, where the true OBD lies one level below the MTD, with an average improvement of 7.9 percentage points, including sizable gains in scenario 3 (57.4\% vs. 40.8\%). In scenarios 5–6, where the OBD coincides with the MTD, BE-BOIN and BE-CRM are comparable. In contrast, all methods exhibit limited performance in scenarios 7–8, where the OBD is two dose levels below the MTD—resulting in correct OBD selection below 22\%, due to stage II’s prioritization of MTD or MTD–1 doses.

One might be disappointed that BE-BOIN offers only limited improvement in correctly identifying the MTD and OBD compared to TITE-BOIN, despite treating, on average, 9.7 more patients. However, this is expected, as backfilling—by design—tends to assign additional patients to lower doses, which may contribute little to accurately identifying the MTD and OBD. The primary objective and benefit of backfilling is to treat more patients within the same timeframe. This not only increases patient access to novel therapies but also allows for the collection of additional data, including toxicity, efficacy, tolerability, pharmacokinetics/pharmacodynamics profiles, biological activity, and biomarkers, thereby providing valuable insights to guide subsequent clinical development decisions. These important benefits are not explicitly demonstrated through the selection probabilities of MTD or OBD.


A major advantage of BE-BOIN over BF-BOIN is its substantially shorter trial duration. Leveraging real-time decision-making in stage I, BE-BOIN reduces trial duration by an average of 15.0 months (range: 13.1–16.9) compared to BF-BOIN’s 64.5 months, while maintaining comparable accuracy and safety. This reduction is primarily achieved during stage I, where BE-BOIN enables dose assignment for new patients with pending data, resulting in a 30–40\% decrease in duration. In contrast, stage II—focused on multiple-dose randomized optimization—follows a fixed schedule, and all designs share the same duration.

\subsection{BE-BOIN vs BE-BOIN without late-onset toxicity}
Table~\ref{tab:BTBF} compares the performance of BE-BOIN and BF-BOIN in the absence of late-onset toxicity. The objective is to evaluate whether ignoring pending backfill patients—as done in BF-BOIN—affects the operating characteristics of the design. Overall, the two designs yield nearly identical performance in terms of dose selection (including MTD identification) and patient allocation at each dose level, whether through the DE or BF components. Trial durations are also comparable. This similarity arises because, at each decision point, all patients in the DE component have complete follow-up, and DE primarily guides dose decisions. These results support the validity of the BF-BOIN approach in settings without late-onset toxicity.


\subsection{Sensitivity Analysis}
In Supplementary material S4, Figure A1 presents sensitivity analysis based on data generated from Log-logistic, Uniform, and Weibull time-to-event distributions. BE-BOIN exhibits consistent performance across all distributions, underscoring its robustness to the underlying time-to-event assumptions. Figure A2 shows results under varying  A/I ratios, with the fixed DLT assessment window of 3 months: A/I = 3 (one patient/month), A/I = 6 (two patients/month), and A/I = 9 (three patients/month). Higher A/I ratios lead to greater reductions in trial duration, more patients treated at the MTD and OBD, and fewer overdosed patients. 
Nonetheless, differences in operating characteristics across A/I settings remain minor. Figure A3 reports results using Weibull-distributed time-to-event data with varying proportions of late-onset toxicities—defined as DLTs occurring in the second half of the assessment window: 30\%, 50\%, and 70\%. As the proportion of late-onset toxicity increases, the number of overdosed patients decreases. This is because a higher proportion of late-onset toxicity results in less data being available for real-time decision-making. BE-BOIN appropriately accounts for this uncertainty by adopting a more conservative dose-escalation strategy, leading to fewer overdosed patients. Notably, the trial duration remains largely unaffected by the timing of DLT onset.

\section{Conclusion}
We propose the BE-BOIN design, which allows backfilling patients into safe and effective doses during dose escalation and accommodates late-onset toxicities. BE-BOIN enables the collection of additional safety and efficacy data to enhance the accuracy and reliability of optimal dose selection and supports real-time dose decisions for new patients. Our simulation studies show that BE-BOIN accurately identifies the MTD/OBD while significantly reducing trial duration. As a byproduct, our results also demonstrate that the BF-BOIN approach of ignoring pending backfilling patients has a negligible impact on the operating characteristics of the design, addressing a common question raised about BF-BOIN.



The BE-BOIN design assumes a binary toxicity endpoint. In some trials, it may be beneficial to consider the grades and types of toxicities, as well as quality of life, to guide dose escalation and selection. One approach is to summarize various grades and types of toxicity using a total toxicity burden score as a composite endpoint \citep{Bekele2004, Lee2012}. The total toxicity burden score can be dichotomized and applied to the BE-BOIN design. Alternatively, we can directly use the total toxicity burden score as the endpoint and utilize a quasi-binomial approach \citep{Yuan2007} or a generalized BOIN approach \citep{Mu2019} to accommodate such continuous or semi-continuous endpoints.

\clearpage
\newpage

\clearpage
\newpage
\definecolor{NoConflict}{RGB}{210,245,210}   
\definecolor{Conflict}{RGB}{255,210,210}     

\begin{table}[ht]
\centering
\caption{Conflicting decisions between the current dose-escalation dose $c$ and a backfilled dose $b$, where $c>b$.}
\label{Dataconflict}
\begin{tabular}{cccc}
\toprule
& \multicolumn{3}{c}{\makecell{\textbf{Decision according to the data}\\\textbf{at current dose }$c$}}\\
\cmidrule(l){2-4}
\makecell[l]{\textbf{Decision according to the}\\\textbf{data at a backfilled dose }$b$}
  & \makecell{Escalate\\($\hat p_c<\lambda_e$)}
  & \makecell{Stay\\($\lambda_e<\hat p_c\le\lambda_d$)}
  & \makecell{De-escalate or \\Eliminate\\($\hat p_c>\lambda_d$)}\\
\midrule
\makecell[c]{Escalate ($\hat p_b<\lambda_e$)}
  & \cellcolor{NoConflict}No conflict
  & \cellcolor{NoConflict}No conflict
  & \cellcolor{NoConflict}No conflict\\
\makecell[c]{Stay ($\lambda_e<\hat p_b\le\lambda_d$)}
  & \cellcolor{Conflict}Conflict
  & \cellcolor{NoConflict}No conflict
  & \cellcolor{NoConflict}No conflict\\
\makecell[c]{De-escalate or Eliminate ($\hat p_b>\lambda_d$)}
  & \cellcolor{Conflict}Conflict
  & \cellcolor{Conflict}Conflict
  & \cellcolor{Conflict}Conflict* \\
\bottomrule
\end{tabular}
*This case does not necessarily mean that $\hat{p}_b > \hat{p}_c$. However, as $\hat{p}_b > \lambda_d$, it means that the additional data from backfilling patients demonstrate alarmingly higher toxicity than what originally observed during the dose escalation (i.e., $\hat{p}_b \le \lambda_e$). Therefore, it is important to reconcile such conflict for patient safety. 
\end{table}

\begin{table}[]
\caption{Operating characteristics of TITE-BOIN, BF-BOIN, BE-BOIN, and BE-CRM. 
}
\label{taboc}
\resizebox{\textwidth}{!}{%
\begin{tabular}{clccccccc}
\toprule
Scenario & Method & \begin{tabular}[c]{@{}c@{}}MTD sel \% \\ \end{tabular}  & OBD sel \% & \begin{tabular}[c]{@{}c@{}}Patients \\ at MTD\end{tabular} & \begin{tabular}[c]{@{}c@{}}Patients \\ at OBD\end{tabular} & \begin{tabular}[c]{@{}c@{}}Patients \\ overdosed\end{tabular} & \begin{tabular}[c]{@{}c@{}}Total \\ Patients\end{tabular} & \begin{tabular}[c]{@{}c@{}}Trial \\ Duration\end{tabular}\\ \hline
1 & TITE-BOIN & 53.9 & 45.9 & 28.6 & 21.6 & 17.8 & 68.0 & 49.1 \\
  & BF-BOIN   & 59.7 & 49.4 & 34.8 & 27.6 & 16.9 & 79.4 & 64.1 \\
  & BE-BOIN   & 56.9 & 47.7 & 32.6 & 26.7 & 17.0 & 76.3 & 47.2 \\
  & BE-CRM    & 53.2 & 43.9 & 33.5 & 23.4 & 18.9 & 75.8 & 45.5 \\
  &  &  &  &  &  &  &  &  \\
2 & TITE-BOIN & 42.7 & 61.9 & 20.7 & 25.7 & 9.7 & 69.6 & 49.9 \\
  & BF-BOIN   & 48.2 & 67.7 & 23.9 & 32.3 & 7.8 & 80.3 & 64.4 \\
  & BE-BOIN   & 47.1 & 65.4 & 22.7 & 30.9 & 7.7 & 77.2 & 48.5 \\
  & BE-CRM    & 45.7 & 63.3 & 24.1 & 29.6 & 9.7 & 76.9 & 47.0 \\
  &  &  &  &  &  &  &  &  \\
3 & TITE-BOIN & 54.1 & 53.8 & 24.8 & 21.4 & 10.7 & 70.0 & 52.4 \\
  & BF-BOIN   & 56.4 & 55.2 & 28.4 & 27.6 & 9.9 & 83.8 & 64.7 \\
  & BE-BOIN   & 57.4 & 57.4 & 26.7 & 26.9 & 9.4 & 80.3 & 51.0 \\
  & BE-CRM    & 44.3 & 40.8 & 28.6 & 20.7 & 16.9 & 80.7 & 50.0 \\
  &  &  &  &  &  &  &  &  \\
4 & TITE-BOIN & 68.3 & 70.9 & 23.5 & 26.9 & 0.0 & 70.0 & 53.6 \\
  & BF-BOIN   & 68.8 & 71.2 & 23.5 & 32.6 & 0.0 & 85.4 & 64.9 \\
  & BE-BOIN   & 67.9 & 70.6 & 23.3 & 31.2 & 0.0 & 81.6 & 51.8 \\
  & BE-CRM    & 86.9 & 67.1 & 31.4 & 30.4 & 0.0 & 81.8 & 51.0 \\
  &  &  &  &  &  &  &  &  \\
5 & TITE-BOIN & 55.5 & 73.1 & 28.5 & 28.5 & 16.4 & 67.6 & 49.0 \\
  & BF-BOIN   & 60.4 & 77.4 & 34.0 & 34.0 & 15.5 & 75.5 & 64.1 \\
  & BE-BOIN   & 58.8 & 76.6 & 32.1 & 32.1 & 15.4 & 73.0 & 47.2 \\
  & BE-CRM    & 56.2 & 76.6 & 33.6 & 33.6 & 16.5 & 72.2 & 45.3 \\
  &  &  &  &  &  &  &  &  \\
6 & TITE-BOIN & 54.1 & 65.5 & 24.8 & 24.8 & 10.7 & 70.0 & 52.4 \\
  & BF-BOIN   & 56.2 & 66.6 & 28.4 & 28.4 & 9.8 & 82.7 & 64.7 \\
  & BE-BOIN   & 56.8 & 65.5 & 26.6 & 26.6 & 9.7 & 79.5 & 51.1 \\
  & BE-CRM    & 44.4 & 67.1 & 28.5 & 28.5 & 16.5 & 79.5 & 50.2 \\
  &  &  &  &  &  &  &  &  \\
7 & TITE-BOIN & 49.5 & 10.5 & 25.0 & 8.3 & 15.4 & 69.9 & 50.8 \\
  & BF-BOIN   & 54.9 & 9.6 & 30.4 & 14.0 & 14.4 & 87.1 & 64.5 \\
  & BE-BOIN   & 51.9 & 10.5 & 28.2 & 13.5 & 13.8 & 82.6 & 49.2 \\
  & BE-CRM    & 47.5 & 9.0 & 28.6 & 10.9 & 19.4 & 83.1 & 48.3 \\
  &  &  &  &  &  &  &  &  \\
8 & TITE-BOIN & 42.8 & 20.5 & 21.9 & 11.4 & 11.2 & 70.0 & 52.1 \\
  & BF-BOIN   & 44.5 & 21.3 & 25.2 & 17.9 & 10.2 & 86.4 & 64.7 \\
  & BE-BOIN   & 46.2 & 22.0 & 23.4 & 16.9 & 9.4 & 82.2 & 50.7 \\
  & BE-CRM    & 37.6 & 15.1 & 26.0 & 13.4 & 15.9 & 82.7 & 49.8 \\ 
\bottomrule
\end{tabular}
}
MTD sel \% denotes the percentage of correct MTD selection after {stage I}, while OBD sel \% denotes the percentage of correct OBD selection after {stage II}.
\end{table}

\begin{table}[h]
\caption{Comparison between BE-BOIN and BF-BOIN design in settings without late-onset toxicity.}
\label{tab:BTBF}
\resizebox{\textwidth}{!}{%
\begin{tabular}{clrcccccc} 
\toprule
Scenario & Method & Metrics & $d_1$ & $d_2$ & $d_3$ & $d_4$ & $d_5$ & Trial Duration \\ 
\midrule
\multirow{4}{*}{1} & \multirow{2}{*}{BF-BOIN} & Sel(\%) & 7.93 & \textbf{58.15} & 28.87 & 4.27 & 0.54 & 17.97 \\
& & Patients & 13.01 & \textbf{15.77} & 7.82 & 1.67 & 0.27 & \\
& \multirow{2}{*}{BE-BOIN} & Sel(\%) & 8.40 & \textbf{57.54} & 29.12 & 4.21 & 0.45 & 18.01 \\
& & Patients & 13.22 & \textbf{15.65} & 7.67 & 1.65 & 0.25 & \\
\\ 
\multirow{4}{*}{2} & \multirow{2}{*}{BF-BOIN} & Sel(\%) & 2.13 & 34.44 & \textbf{45.89} & 15.71 & 1.80 & 18.08 \\
& & Patients & 8.45 & 14.27 & \textbf{10.60} & 3.96 & 0.80 & \\
& \multirow{2}{*}{BE-BOIN} & Sel(\%) & 2.16 & 34.99 & \textbf{45.66} & 15.35 & 1.82 & 18.09 \\
& & Patients & 8.43 & 14.39 & \textbf{10.53} & 3.91 & 0.76 & \\
\\ 
\multirow{4}{*}{3} & \multirow{2}{*}{BF-BOIN} & Sel(\%) & 0.04 & 1.88 & 16.47 & \textbf{54.94} & 26.67 & 18.21 \\
& & Patients & 4.72 & 7.58 & 11.74 & \textbf{11.08} & 4.88 & \\
& \multirow{2}{*}{BE-BOIN} & Sel(\%) & 0.08 & 2.05 & 15.87 & \textbf{55.92} & 26.08 & 18.22 \\
& & Patients & 4.71 & 7.59 & 11.80 & \textbf{11.08} & 4.90 & \\
\\ 
\multirow{4}{*}{4} & \multirow{2}{*}{BF-BOIN} & Sel(\%) & 0.00 & 0.18 & 2.57 & 27.23 & \textbf{70.02} & 18.30 \\
& & Patients & 4.03 & 5.95 & 9.30 & 12.12 & \textbf{9.81} & \\
& \multirow{2}{*}{BE-BOIN} & Sel(\%) & 0.00 & 0.26 & 2.44 & 27.23 & \textbf{70.07} & 18.32 \\
& & Patients & 4.02 & 5.90 & 9.22 & 12.14 & \textbf{9.86} & \\
\\ 
\multirow{4}{*}{5} & \multirow{2}{*}{BF-BOIN} & Sel(\%) & 10.81 & \textbf{59.58} & 25.77 & 3.26 & 0.33 & 17.99 \\
& & Patients & 11.36 & \textbf{15.40} & 7.07 & 1.46 & 0.20 & \\
& \multirow{2}{*}{BE-BOIN} & Sel(\%) & 10.96 & \textbf{59.67} & 25.27 & 3.34 & 0.50 & 17.97 \\
& & Patients & 11.45 & \textbf{15.27} & 7.04 & 1.47 & 0.22 & \\
\\ 
\multirow{4}{*}{6} & \multirow{2}{*}{BF-BOIN} & Sel(\%) & 0.08 & 2.04 & 16.00 & \textbf{55.55} & 26.33 & 18.22 \\
& & Patients & 4.71 & 7.59 & 10.91 & \textbf{11.01} & 4.92 & \\
& \multirow{2}{*}{BE-BOIN} & Sel(\%) & 0.04 & 2.09 & 16.01 & \textbf{55.35} & 26.51 & 18.20 \\
& & Patients & 4.70 & 7.56 & 10.88 & \textbf{10.94} & 4.87 & \\
\\ 
\multirow{4}{*}{7} & \multirow{2}{*}{BF-BOIN} & Sel(\%) & 0.51 & 14.69 & \textbf{52.86} & 27.15 & 4.77 & 18.13 \\
& & Patients & 9.77 & 13.25 & \textbf{12.64} & 5.96 & 1.39 & \\
& \multirow{2}{*}{BE-BOIN} & Sel(\%) & 0.51 & 15.10 & \textbf{52.87} & 26.35 & 5.14 & 18.12 \\
& & Patients & 9.83 & 13.25 & \textbf{12.61} & 5.97 & 1.42 & \\
\\ 
\multirow{4}{*}{8} & \multirow{2}{*}{BF-BOIN} & Sel(\%) & 0.05 & 2.94 & 25.10 & \textbf{42.82} & 29.09 & 18.21 \\
& & Patients & 4.54 & 10.44 & 12.30 & \textbf{9.80} & 4.80 & \\
& \multirow{2}{*}{BE-BOIN} & Sel(\%) & 0.04 & 3.16 & 24.20 & \textbf{44.15} & 28.45 & 18.20 \\
& & Patients & 4.55 & 10.44 & 12.23 & \textbf{9.83} & 4.84 & \\
\bottomrule
\end{tabular}%
}
sel \% denotes the percentage of selecting each dose as the MTD after {stage I}. Patients represent the average number of patients treated at each dose.
\end{table}

 \clearpage
 \newpage

\begin{center}

{\Large \bf Supplementary Materials}
   
\end{center}
   
\setcounter{section}{0}
\section{Decision table}
We utilize the same TITE model and suspension rules as in the TITE-BOIN design, allowing the decision table to be directly applied in the BE-BOIN framework. During the trial, at each decision point, the following data are recorded for each dose: the total number of patients, the number of observed DLTs, the number of pending patients, and two key metrics—TF (Standardized Total Follow-Up Time) and MF (Standardized Minimum Follow-Up Time). TF represents the cumulative follow-up time of all pending patients divided by the DLT assessment window, while MF is calculated as the minimum follow-up time among all pending patients divided by the DLT assessment window. These metrics ensure consistency and reliability in decision-making.

\begin{table}[ht]
\centering
\caption{Dose-escalation and de-escalation boundaries up to 9 patients, with a target DLT rate of 0.25 and cohort size of 3. The cutoff A in suspension rule 1 is 51 \%, and the cutoff B in suspension rule 2 is 25 \%.}
\label{tab:decision_table}
\resizebox{\textwidth}{!}{%
\begin{tabular}{cccccccc}
\hline\hline
\textbf{No. patients} & \textbf{No. DLT} & \textbf{No. pending} & \textbf{Suspension} & \textbf{Escalation} & \textbf{Stay} & \textbf{De-escalation} \\ \hline
3 & 0 & 0 & No & Yes & No & No \\
3 & 0 & 1 & MF $<$ 0.25 & MF $\geq$ 0.25 & No & No \\
3 & 0 & $\geq$ 2 & Yes & No & No & No \\
3 & 1, 2 & $\leq$ 2 & No & No & No & Yes \\
3 & 3 & 0 & No & No & No & Yes \& Eliminate \\ \hline
6 & 0 & 0 & No & Yes & No & No \\
6 & 0 & 1, 2 & MF $<$ 0.25 & MF $\geq$ 0.25 & No & No \\
6 & 0 & $\geq$ 3 & Yes & No & No & No \\
6 & 1 & 0 & No & Yes & No & No \\
6 & 1 & 1 & MF $<$ 0.25 \& TF $\geq$ 0.22 & MF $\geq$ 0.25 \& TF $\geq$ 0.22 & TF $<$ 0.22 & No \\
6 & 1 & 2 & MF $<$ 0.25 \& TF $\geq$ 1.38 & MF $\geq$ 0.25 \& TF $\geq$ 1.38 & TF $<$ 1.38 & No \\
6 & 1 & $\geq$ 3 & Yes & No & No & No \\
6 & 2, 3 & $\leq$ 4 & No & No & No & Yes \\
6 & $\geq$ 4 & $\leq$ 2 & No & No & No & Yes \& Eliminate \\ \hline
9 & 0 & 0 & No & Yes & No & No \\
9 & 0 & 1–4 & MF $<$ 0.25 & MF $\geq$ 0.25 & No & No \\
9 & 0 & $\geq$ 5 & Yes & No & No & No \\
9 & 1 & 0 & No & Yes & No & No \\
9 & 1 & 1–3 & MF $<$ 0.25 & MF $\geq$ 0.25 & No & No \\
9 & 1 & 4 & MF $<$ 0.25 \& TF $\geq$ 0.66 & MF $\geq$ 0.25 \& TF $\geq$ 0.66 & TF $<$ 0.66 & No \\
9 & 1 & $\geq$ 5 & Yes & No & No & No \\
9 & 2 & $\leq$ 4 & No & No & Yes & No \\
9 & 2 & $\geq$ 5 & Yes & No & No & No \\
9 & 3, 4 & $\leq$ 6 & No & No & No & Yes \\
9 & $\geq$ 5 & $\leq$ 4 & No & No & No & Yes \& Eliminate \\ \hline\hline
\end{tabular}%
}
\footnotesize
\textit{Note: “No. patients” is the total number of patients treated at one dose. “No. DLT” is the number of patients who experienced DLT at one dose. “No. pending” denotes the number of patients whose DLT data are pending. TF: standardized total follow-up time of pending patients. MF: standardized minimum follow-up time of pending patients. When a dose is eliminated, all higher doses should also be eliminated.}
\end{table}
\clearpage

\section{True scenarios}
\begin{table}[ht]
\centering
\caption{True toxicity ($p_T$) and efficacy ($p_E$)  across dose levels.}
\label{tab:scenario}
\begin{tabular}{ccccccc}
\hline\hline
Scenario & True Rate & $d_1$ & $d_2$ & $d_3$ & $d_4$ & $d_5$ \\ 
\hline
1 & $p_T$ & 0.10 & 0.18 & 0.35 & 0.40 & 0.50 \\ 
 & $p_E$ & 0.35 & 0.35 & 0.37 & 0.39 & 0.39 \\
\\
2 & $p_T$ & 0.05 & 0.15 & 0.25 & 0.35 & 0.50 \\ 
 & $p_E$ & 0.10 & 0.35 & 0.35 & 0.38 & 0.39 \\
\\
3 & $p_T$ & 0.02 & 0.06 & 0.10 & 0.20 & 0.35 \\ 
 & $p_E$ & 0.05 & 0.10 & 0.35 & 0.35 & 0.40 \\
\\
4 & $p_T$ & 0.01 & 0.03 & 0.05 & 0.12 & 0.22 \\ 
 & $p_E$ & 0.05 & 0.10 & 0.15 & 0.35 & 0.36 \\
\\
5 & $p_T$ & 0.10 & 0.20 & 0.35 & 0.43 & 0.50 \\ 
 & $p_E$ & 0.10 & 0.36 & 0.37 & 0.40 & 0.41 \\
\\
6 & $p_T$ & 0.02 & 0.06 & 0.10 & 0.20 & 0.35 \\ 
 & $p_E$ & 0.05 & 0.10 & 0.15 & 0.35 & 0.37 \\
\\
7 & $p_T$ & 0.05 & 0.10 & 0.20 & 0.35 & 0.40 \\ 
 & $p_E$ & 0.35 & 0.36 & 0.37 & 0.40 & 0.41 \\
\\
8 & $p_T$ & 0.01 & 0.05 & 0.15 & 0.18 & 0.35 \\ 
 & $p_E$ & 0.05 & 0.35 & 0.36 & 0.37 & 0.38 \\
\hline\hline
\end{tabular}
\end{table}
\clearpage

\section{Model setting for BE-CRM design}
\citet{BFCRM} employ a two-parameter logistic model, defined as:
$$
p_j = \frac{\exp{(a_0 + \beta d^*_j)}}{1 + \exp{(a_0 + \beta d^*_j)}}, \quad j = 1, \dots, J,
$$
where $d^*_j$ represents the standardized dose at level $j$. In the simulations, we adjusted $d^*_j$ to set the prior toxicity rate skeleton as (0.01, 0.10, 0.25, 0.45, 0.60), positioning the target DLT rate in the middle. The prior of $a_0$ is $N(\log(16),4)$, and prior of $\beta$ is $N(0, 1)$.

Following the methodology described in the original paper \citep{BFCRM}, two early stop rules are implemented for safety. The first rule stops the trial if the lowest dose is deemed too toxic, defined by:
$$
\Pr(p_1 > \phi_T | D) > \eta,
$$
where $\eta$ is set at 0.80 for the simulations. The second rule excludes a dose, and all higher doses, from further experimentation if the probability that the toxicity exceeds the target toxicity is high, specified as:
$$
\Pr(p_j > \phi_T|D) > 0.95.
$$
This second safety criterion relies on a beta-binomial model with a Beta(1,1) prior.

At the end of trial, the same logistic model is fitted based on all cumulative data. And the MTD is selected from the dose having the toxicity rate closest to target DLT rate. 

\clearpage
\section{Results of sensitivity analysis}
\begin{figure}[ht]
\centering
\hspace*{-1.5cm}  
\includegraphics[width=1\textwidth]{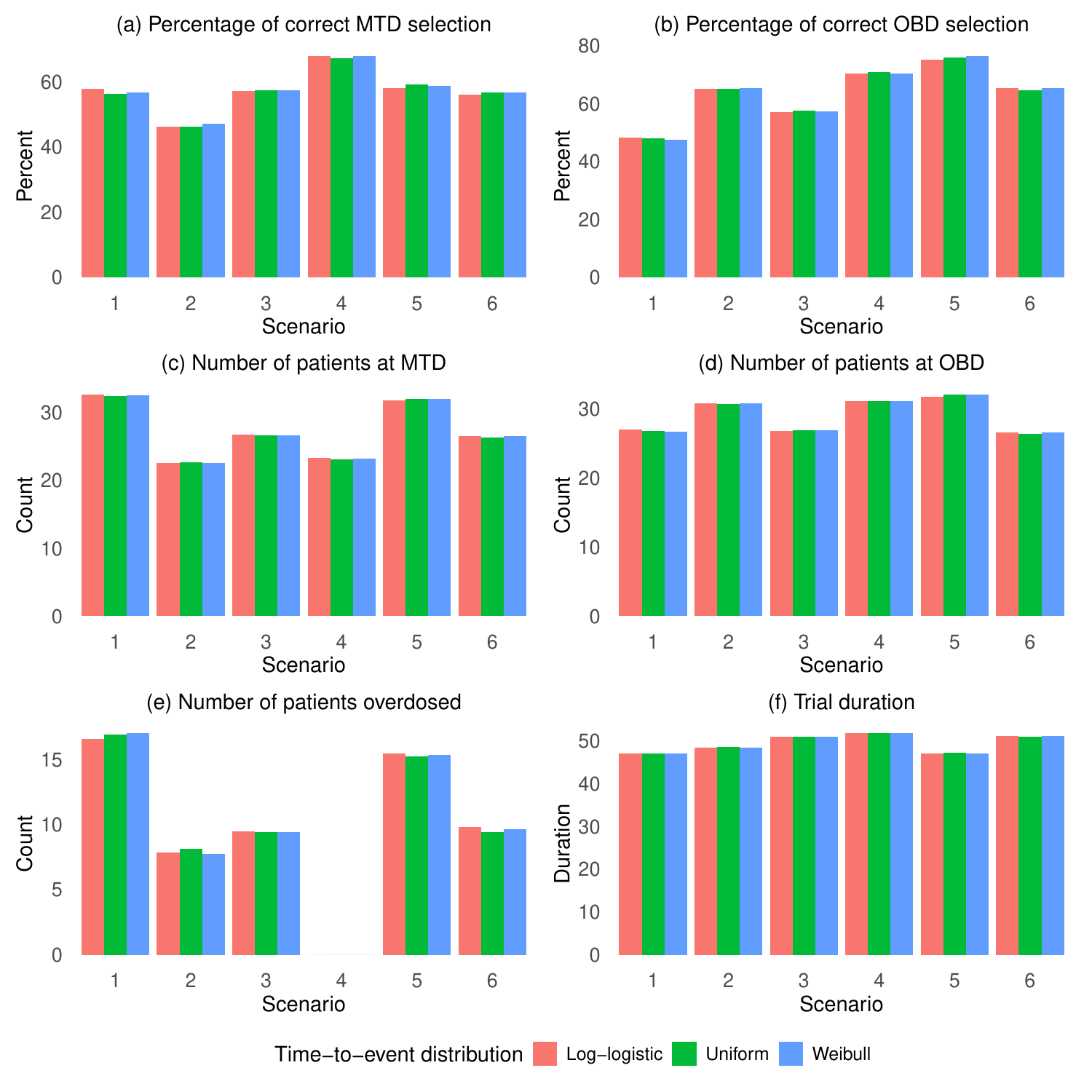}
    \caption{Sensitivity analysis of BE-BOIN design under different time-to-event distribution, including Weibull, Uniform and Log-logistic distribution.}
    \label{Sens:timeToEvent}
\end{figure}

\begin{figure}
\centering
\hspace*{-1.5cm}  
\includegraphics[width=1\textwidth]{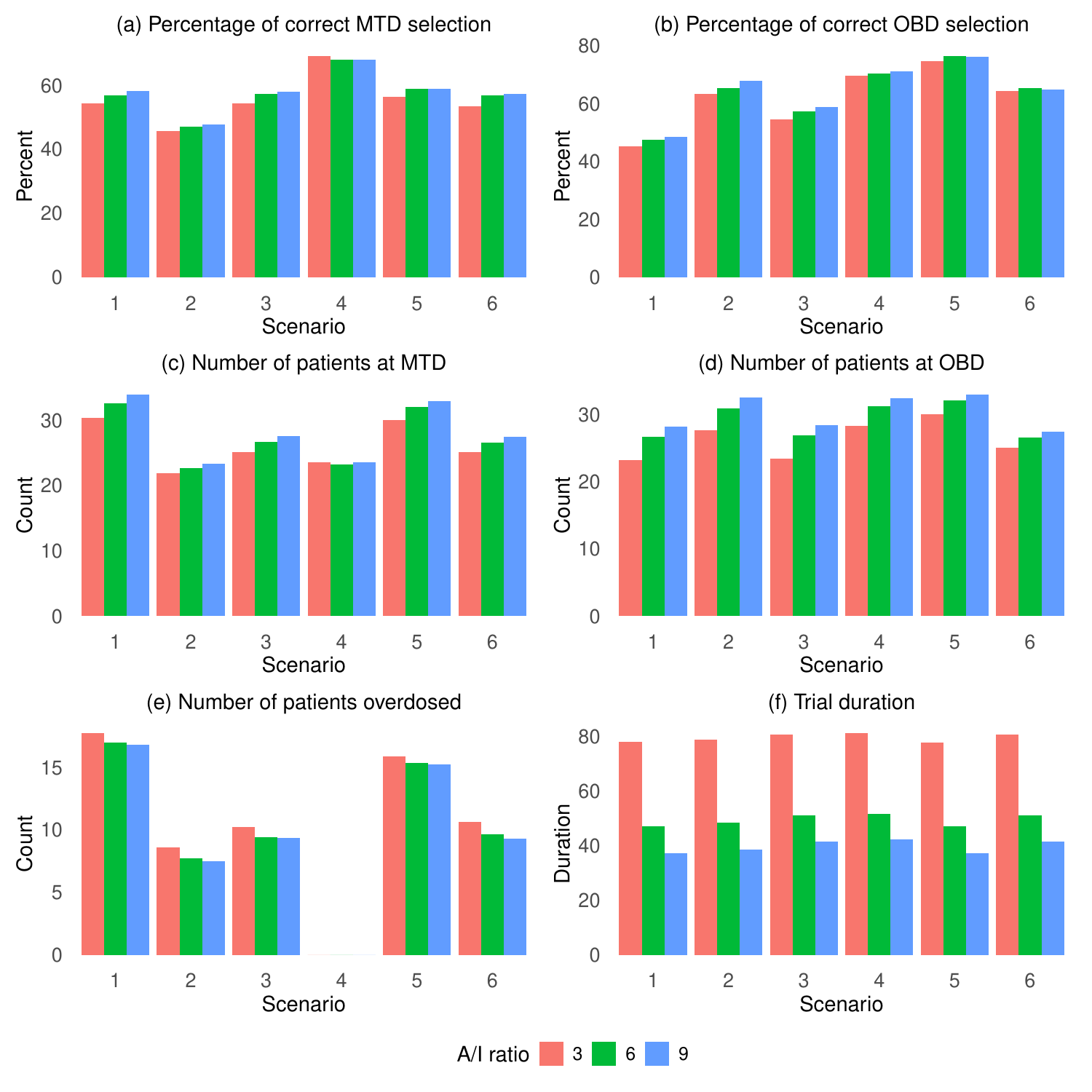}
    \caption{Sensitivity analysis of BE-BOIN design under different A/I ratios.}
    \label{Sens:AIratio}
\end{figure}

\begin{figure}
\centering
\hspace*{-1.5cm}  
\includegraphics[width=1\textwidth]{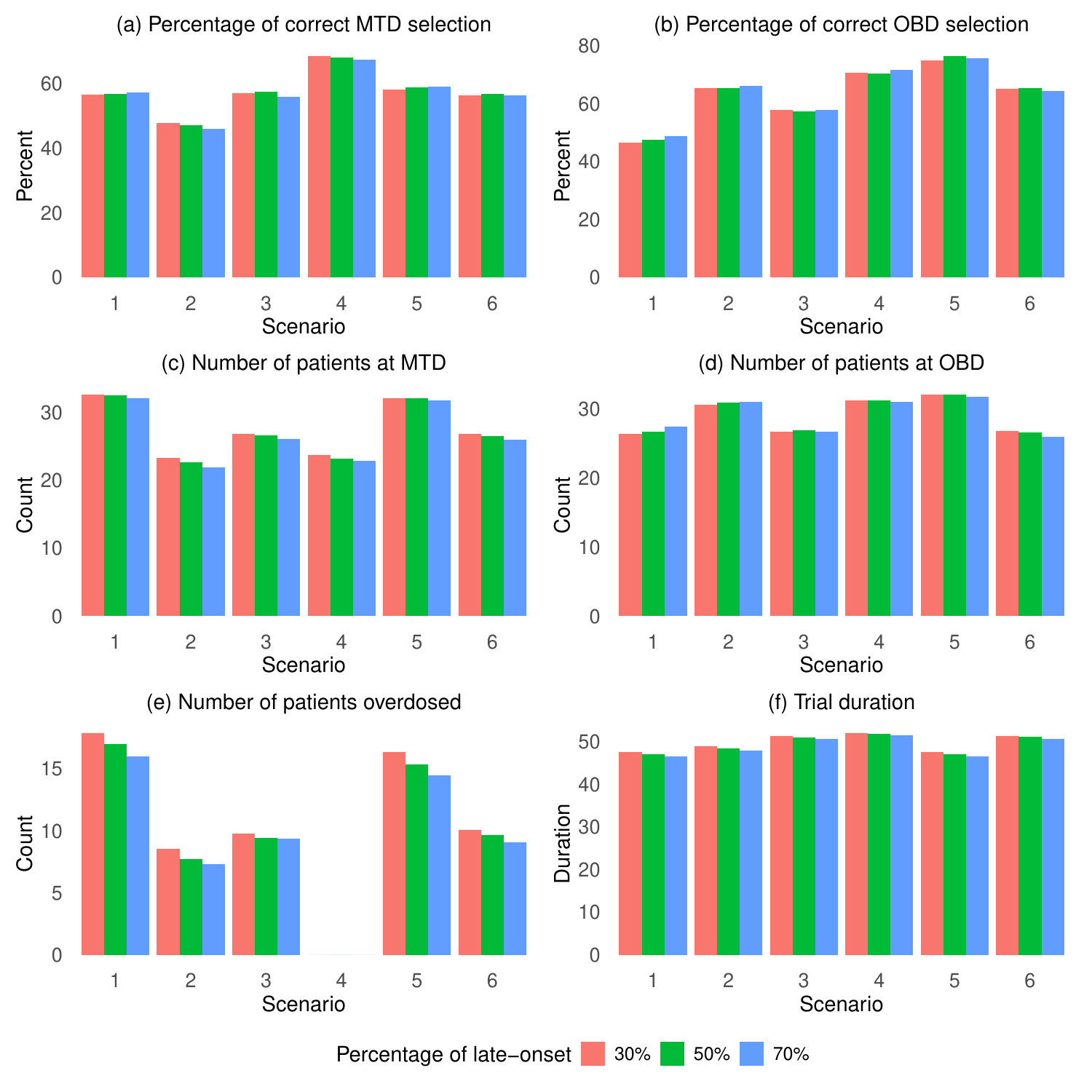}
    \caption{Sensitivity analysis of BE-BOIN design under different late-onset profiles for toxicity.}
    \label{Sens:lateOnset}
\end{figure}

\clearpage


\clearpage
\newpage
\begin{thebibliography}{99}
\bibitem[Shah et al., 2021]{shah2021drug}Shah M, Rahman A, Theoret MR, Pazdur R. The Drug-Dosing Conundrum in Oncology - When Less Is More. {\it N Engl J Med.} 2021;385(16):1445-1447. 

\bibitem[Fourie Zirkelbach et al., 2022]{Zirkelbach2022}Fourie Zirkelbach J, Shah M, Vallejo J, et al. Improving Dose-Optimization Processes Used in Oncology Drug Development to Minimize Toxicity and Maximize Benefit to Patients. {\it J Clin Oncol.} 2022;40(30):3489-3500.

\bibitem[Bekele \& Thall., 2004]{Bekele2004}Bekele BN, and Thall PF. (2004) Dose-finding based on multiple toxicities in a soft tissue sarcoma trial. {\it J. Am. Statist. Ass.}, 99, 26–35.

\bibitem[Lee et al., 2012]{Lee2012}Lee SM, Hershman DL, Martin P, Leonard JP, Cheung YK. Toxicity burden score: a novel approach to summarize multiple toxic effects. {\it Ann Oncol.} 2012;23(2):537-541.

\bibitem[Yuan et al., 2007]{Yuan2007}Yuan Z, Chappell R, Bailey H. The continual reassessment method for multiple toxicity grades: a Bayesian quasi-likelihood approach. {\it Biometrics.} 2007;63(1):173-179. 

\bibitem[Mu et al., 2019]{Mu2019} Mu R, Yuan Y, Xu J, Mandrekar SJ, and Yin J. (2019). gBOIN: A unified model-assisted phase I trial design accounting for toxicity grades, and binary or continuous end points. {\it J R Stat Soc Ser C Appl Stat}, 68(2), 289–308.

\bibitem[Postel-Vinay et al., 2011]{Postel2011} Postel-Vinay S, Gomez-Roca C, Molife LR, et al. Phase I trials of molecularly targeted agents: should we pay more attention to late toxicities? {\it J Clin Oncol}. 2011; 29(13): 1728- 1735.

\bibitem[Weber et al., 2015]{Weber2015} Weber JS, Yang JC, Atkins MB, Disis ML. Toxicities of Immunotherapy for the Practitioner. {\it J Clin Oncol}. 2015;33(18):2092-2099.

\bibitem[Kanjanapan et al., 2019]{Kanjanapan2019} Kanjanapan Y, Day D, Butler MO, et al. Delayed immune-related adverse events in assessment for dose-limiting toxicity in early phase immunotherapy trials. {\it Eur J Cancer}. 2019;107:1-7.

\bibitem[Zhou et al., 2018]{Zhou2018}Zhou H, Yuan Y, Nie L. Accuracy, Safety, and Reliability of Novel Phase I Trial Designs. {\it Clin Cancer Res}. 2018;24(18):4357-4364. 

\bibitem[FDA, 2022]{FD2022}US Food and Drug Administration guidance. Optimizing the Dosage of Human Prescription Drugs and Biological Products for the Treatment of Oncologic Diseases Guidance for Industry,  

\bibitem[Huang \& Yuan, 2023]{Huang23} Huang B, Yuan, Y (2023). Design strategies for dose optimization in oncology drug development. {\it American Statistical Association Biopharmaceutical Report} 30 (No. 2), 27-32.


\bibitem[Barnett et al., 2023]{BFCRM} Barnett H, Boix O, Kontos D, Jaki T. Backfilling cohorts in phase I dose-escalation studies. {\it Clin Trials}. 2023;20(3):261-268.

\bibitem[Dehbi et al., 2021]{BF2021} Dehbi HM, O'Quigley J, Iasonos A. Controlled backfill in oncology dose-finding trials. {\it Contemp Clin Trials}. 2021;111:106605.

\bibitem[Zhao et al., 2024]{BFBOIN} Zhao Y, Yuan Y, Korn EL, Freidlin B. Backfilling Patients in Phase I Dose-Escalation Trials Using Bayesian Optimal Interval Design (BOIN). {\it Clin Cancer Res}. 2024;30(4):673-679.

\bibitem[Zhao et al., 2024]{BARD}Zhao Y, Liu R, Lin J, Yuan Y. BARD: A seamless two-stage dose optimization design integrating backfill and adaptive. arXiv, https://doi.org/10.48550/arXiv.2409.15663.

\bibitem[Yuan \& Yin, 2011]{EMCRM}Yuan Y, Yin G. Robust EM Continual Reassessment Method in Oncology Dose Finding. {\it J Am Stat Assoc}. 2011;106(495):818-831.

\bibitem[Jiang \& Yuan, 2023]{Jiang23} Jiang L, Yuan Y. Seamless phase II/III design: a useful strategy to reduce the sample size for dose optimization. {\it J Natl Cancer Inst.} 115(9):1092-1098.


\bibitem[Liu and Yuan, 2015]{BOIN2015}Liu S, Yuan Y. Bayesian optimal interval designs for phase I clinical trials. {\it J R Stat Soc Ser C Appl Stat}. 2015;64:507–23. 

\bibitem[Yuan et al., 2016]{BOIN2016}Yuan Y, Hess KR, Hilsenbeck SG, Gilbert MR. Bayesian optimal interval design: a simple and well-performing design for phase I oncology trials. {\it Clin Cancer Res}. 2016;22:4291–301.

\bibitem[Yuan et al., 2018]{TITEBOIN}Yuan Y, Lin R, Li D, Nie L, Warren KE. Time-to-Event Bayesian Optimal Interval Design to Accelerate Phase I Trials. {\it Clin Cancer Res}. 2018;24(20):4921-4930.

\bibitem[Chen et al., 2024]{BFQoL} Chen X, Zhang J, Li B, Yan F. Determining doses for backfill cohorts based on patient-reported outcome. {\it BMC Med Res Methodol}. 2024;24(1):270.

\bibitem[O'Quigley et al., 1990]{CRM1990}O'Quigley J, Pepe M, Fisher L. Continual reassessment method: a practical design for phase I clinical trials in cancer. {\it Biometrics}. 1990;46:33–48.

\bibitem[Cheung and Chappell, 2000]{TITECRM}Cheung YK, Chappell R. Sequential designs for phase I clinical trials with late-onset toxicities. {\it Biometrics}. 2000;56(4):1177-1182.

\bibitem[Liu  et al., 2013]{DACRM}Liu S, Yin G, Yuan Y. BAYESIAN DATA AUGMENTATION DOSE FINDING WITH CONTINUAL REASSESSMENT METHOD AND DELAYED TOXICITY. {\it Ann Appl Stat}. 2013; 7(4): 1837-2457.

\bibitem[Yuan et al., 2018]{TITEBOIN} Yuan Y, Lin R, Li D, Nie L, Warren KE. Time-to-Event Bayesian Optimal Interval Design to Accelerate Phase I Trials. {\it Clin Cancer Res}. 2018;24(20):4921-4930.

\bibitem[Lin and Yuan, 2020]{TITEKeyboard} Lin R, Yuan Y. Time-to-event model-assisted designs for dose-finding trials with delayed toxicity. {\it Biostatistics}. 2020;21(4):807-824.

\bibitem[Chen et al., 2025]{TITEBOIN2} Chen K, Chen T, Zhang Y, Lin R, Yuan Y. Practical Considerations for Using the TITE-BOIN Design to Handle Late- onset Toxicity or Fast Accrual in Phase I Trials. {\it Clin Cancer Res} doi: 10.1158/1078-0432.CCR-24-3669. 

\bibitem[Yuan et al., 2022]{Yuanbook2022}Yuan Y, Lin R, Lee JJ. Model-assisted Bayesian designs for dose finding and optimization: methods and applications. {\it Boca Raton, FL: Chapman and Hall/CRC}, 2022

\bibitem[Biard et al., 2024]{Biard2024} Biard L, Andrillon A, Silva RB, Lee SM. Dose optimization for cancer treatments with considerations for late-onset toxicities. {\it Clin Trials}. 2024;21(3):322-330.

\bibitem[Yuan et al., 2024]{Yuan2024DOreview}Yuan Y, Zhou H, Liu S. Statistical and practical considerations in planning and conduct of dose-optimization trials. {\it Clin Trials}. 2024;21(3):273-286.



\bibitem[Pin et al., 2024]{Pin2024}Pin L, Villar SS, Dehbi HM. Implementing and assessing Bayesian response-adaptive randomisation for backfilling in dose-finding trials. {\it Contemp Clin Trials}. 2024;142:107567.


\bibitem[Yang et al., 2024]{Yang24MERIT}Yang P, Li D, Lin R, Huang B, Yuan Y. Design and sample size determination for multiple-dose randomized phase II trials for dose optimization. {\it Stat Med}. 2024;43(15):2972-2986.

\bibitem[Zhou et al., 2019]{UBOIN}Zhou Y, Lee JJ, Yuan Y. A utility-based Bayesian optimal interval (U-BOIN) phase I/II design to identify the optimal biological dose for targeted and immune therapies. Stat Med. 2019;38(28):5299-5316.

\bibitem[Lin et al., 2020]{BOIN12}Lin R, Zhou Y, Yan F, Li D, Yuan Y. BOIN12: Bayesian Optimal Interval Phase I/II Trial Design for Utility-Based Dose Finding in Immunotherapy and Targeted Therapies. {\it JCO Precis Oncol}. 2020;4:PO.20.00257.

\bibitem[D'Angelo et al., 2024]{UMET}D'Angelo G, Gong M, Marshall J, Yuan Y, Li X. U-MET: Utility-based dose optimization approach for multiple-dose randomized trial designs. {\it Stat Biopharm Res.} 2024;17(2):211-221.




\end{thebibliography}
\end{document}